\documentclass{PoS}

\usepackage{citesort}
\usepackage{graphicx}
\usepackage{bm}

\title{Anisotropic hydrodynamics -- basic concepts}

\ShortTitle{Anisotropic hydrodynamics}

\author{\speaker{Wojciech Florkowski}
\\
Institute of Physics, Jan Kochanowski University, ul. Swietokrzyska 15, PL-25406 Kielce, Poland and The H. Niewodniczanski Institute of Nuclear Physics, Polish Academy of Sciences, PL-31342 Krakow, Poland\\
E-mail: \email{Wojciech.Florkowski@ifj.edu.pl}} 
       
\author{{Mauricio Martinez} \\
Departamento de F\'isica de Part\'iculas and IGFAE, Universidade de Santiago de Compostela, E-15782 Santiago de Compostela, Galicia, Spain\\
E-mail: \email{mauricio.martinez@usc.es}} 

\author{{Radoslaw Ryblewski}  \\
The H. Niewodniczanski Institute of Nuclear Physics, Polish Academy of Sciences, PL-31342 Krakow, Poland\\
E-mail: \email{Radoslaw.Ryblewski@ifj.edu.pl}} 
       
\author{{Michael Strickland}  \\
Department of Physics, Kent State University, Kent, OH 44242 United States \\
and Frankfurt Institute for Advanced Studies, Ruth-Moufang-Strasse 1, D-60438, Frankfurt am Main, Germany \\
E-mail: \email{mstrick6@kent.edu}} 

\abstract{Due to the rapid longitudinal expansion of the quark-gluon plasma created in relativistic heavy ion collisions, potentially large local rest frame momentum-space anisotropies are generated.  The magnitude of these momentum-space anisotropies can be so large as to violate the central assumption of canonical viscous hydrodynamical treatments which linearize around an isotropic background.  In order to better describe the early-time dynamics of the quark gluon plasma, one can consider instead expanding around a locally anisotropic background which results in a dynamical framework called anisotropic hydrodynamics.  In this proceedings contribution we review the basic concepts of the anisotropic hydrodynamics framework presenting viewpoints from both the phenomenological and microscopic points of view.}

\FullConference{Xth Quark Confinement and the Hadron Spectrum,\\
		October 8-12, 2012\\
		TUM Campus Garching, Munich, Germany}

\begin{document}

\section{Introduction}

The space-time evolution of matter created in relativistic heavy-ion collisions seems to be very well described by relativistic viscous hydrodynamics \cite{Israel:1979wp,Muronga:2003ta,Baier:2006um,Romatschke:2007mq,Dusling:2007gi,Luzum:2008cw,Song:2008hj,Denicol:2010tr,Schenke:2011tv,Shen:2011eg,Bozek:2011wa,Niemi:2011ix,Bozek:2012qs}.  Nevertheless, in the early stages of the collisions,  due to large flow gradients the viscous corrections to the perfect-fluid energy-momentum tensor become very large and the system becomes highly anisotropic in momentum space (in the local rest frame). Typically, the transverse pressure, $P_\perp$, is much larger than the longitudinal pressure, $P_\parallel$, where the directions are specified with respect to the beam axis. Generally speaking, large momentum-space anisotropies pose a problem for 2nd-order viscous hydrodynamics, since it relies on a linearization around an isotropic background. It has been shown that large linear corrections
  generate unphysical results such as negative particle pressures, negative one-particle distribution functions, etc.~\cite{Martinez:2009mf}. 

A similar aspect of the same problem is that microscopic models of the early stages of relativistic heavy-ion collisions suggest that the produced system is highly anisotropic \cite{Kovchegov:2009he}. In the theory of the Color Glass Condensate (CGC), at very early proper-times, $\tau \ll 1/Q_s$, where $Q_s$ is the saturation scale, the classical gluon fields lead to an energy-momentum tensor which has the form $T^{\mu\nu} = \hbox{diag}(\varepsilon,\varepsilon,\varepsilon,-\varepsilon)$ \cite{Lappi:2006hq,Fukushima:2007ja}. This form implies a negative value of the longitudinal pressure with the transverse pressure equal to the energy density. At later proper times, $\tau \gg 1/Q_s$, both the analytical perturbative approaches \cite{Kovchegov:2005ss} and the full numerical simulations \cite{Krasnitz:2002mn} lead to the form $T^{\mu\nu} = \hbox{diag}(\varepsilon,\varepsilon/2,\varepsilon/2,0)$. This form implies that the longitudinal pressure is zero. Consequently, proper matching of the results of the microscopic models with the hydrodynamic description (where the energy-momentum tensor should be close to the isotropic form) is not straightforward.

In addition to the issues faced at early times, one finds that as one approaches the transverse edge of the interaction region, corrections to ideal hydrodynamics can be large at all times.  As a result, one expects large momentum-space anisotropies to be present at the edges \cite{Martinez:2009mf} and, as a result, one is motivated to find better approximation schemes to describe the dynamics in these regions.  Finally, and perhaps most importantly, physically one expects entropy production to vanish in two limits:  the ideal hydrodynamical limit (vanishing shear viscosity) and the free streaming limit (infinite shear viscosity); however, within viscous hydrodynamics, entropy production is a monotonically increasing function of the shear viscosity.  In the large shear viscosity limit, as one might expect, viscous hydrodynamics becomes a poor approximation and one is, once again, motivated to find an alternative framework.

The difficulties eluded to above motivated the development of a reorganization of viscous hydrodynamics in which one incorporates the possibility of large momentum-space anisotropies at leading order. The inclusion of large anisotropies also allows for direct matching with the theories such as CGC. This framework has been dubbed {\it anisotropic hydrodynamics} \cite{Florkowski:2010cf,Martinez:2010sc,Ryblewski:2010bs,Martinez:2010sd,Ryblewski:2011aq,Martinez:2012tu,Ryblewski:2012rr}.  Below we will briefly review the basic ideas underpinning the framework and direct the reader to various references for more details. A more self-contained presentation of the concepts of anisotropic hydrodynamics can be found, e.g.,  in Ref.~\cite{Ryblewski:2012it}.

\section{Anisotropic hydrodynamics}

The framework of anisotropic hydrodynamics may be introduced in (at least) two ways. The first method is phenomenological and refers directly to tensor structures describing an anisotropic fluid \cite{Florkowski:2010cf,Ryblewski:2010bs,Ryblewski:2011aq,Ryblewski:2012rr}. The second method is microscopic; one uses the Boltzman equation for the parton (gluon) distribution function and makes an expansion around the anisotropic background \cite{Martinez:2010sc,Martinez:2010sd,Martinez:2012tu}. In this Section we outline the first method. The microscopic approach will be discussed in the next Section.

The dynamic equations of anisotropic hydrodynamics follow from the equations
\begin{eqnarray}
\partial_\mu T^{\mu \nu} &=& 0 \, ,  \\
\partial_\mu \sigma^{\mu} &= &\Sigma \, , 
\label{enmomcon}
\end{eqnarray}
which express the energy-momentum conservation and entropy production laws, respectively. The energy-momentum tensor $T^{\mu \nu}$ has the form
\begin{eqnarray}
T^{\mu \nu} = \left( \varepsilon  + P_{\perp}\right) U^{\mu}U^{\nu} - P_{\perp} \, g^{\mu\nu} - (P_{\perp} - P_{\parallel}) V^{\mu}V^{\nu}.
\label{Taniso}
\end{eqnarray}
We note that Eq.~(\ref{Taniso}) reproduces the energy-momentum tensor of the perfect fluid if the two pressures are equal. Similarly, in the case $\Sigma=0$ the entropy production law in (\ref{enmomcon}) is reduced to the entropy conservation law. The four-vector $U^{\mu}$ describes the flow of matter and $V^{\mu}$ defines the beam ($z$) axis. In the general (3+1)-dimensional case, we use the following parameterizations: $U^\mu = (u_0 \cosh \vartheta, u_x, u_y, u_0 \sinh \vartheta)$ and $V^\mu = (	 \sinh \vartheta, 0, 0,  \cosh \vartheta)$, where $u_x$ and $u_y$ are the transverse components of the four-velocity field and $\vartheta$ is the longitudinal fluid rapidity. The entropy flux $\sigma^{\mu}$ equals $\sigma \, U^\mu$, where $\sigma$ is the non-equilibrium entropy density.

One can show \cite{Florkowski:2010cf} that instead of $P_{\parallel}$ and $P_{\perp}$ it is more convenient to use the entropy density $\sigma$ and the {\it anisotropy parameter} $x$ as two independent variables (one may use the approximation $P_{\parallel}/P_{\perp} \approx x^{-3/4}$). Similarly to standard hydrodynamics with vanishing baryon chemical potential, the energy density $\varepsilon$, the entropy density  $\sigma$, and the anisotropy parameter $x$ are related through the \textit{generalized} equation of state  \cite{Ryblewski:2010bs}
\begin{eqnarray}
\varepsilon (x,\sigma)&=&  \varepsilon_{\rm qgp}(\sigma) r(x), \label{epsilon2b}  \\ \nonumber 
P_\perp (x,\sigma)&=&  P_{\rm qgp}(\sigma) \left[r(x) + 3 x r^\prime(x) \right], \label{PT2b}   \\ \nonumber 
P_\parallel (x,\sigma)&=&  P_{\rm qgp}(\sigma) \left[r(x) - 6 x r^\prime(x) \right] , \label{PL2b} 
\end{eqnarray}
where the functions $\varepsilon_{\rm qgp}$ and  $P_{\rm qgp}$ correspond to a realistic {\em isotropic} QGP equation of state, see e.g. Ref. \cite{Chojnacki:2007jc}. The function $r(x)$ characterizes  properties of the fluid which exhibits the pressure anisotropy specified by the variable $x$ \cite{Rebhan:2008uj,Martinez:2008di,Florkowski:2010cf}
\begin{equation}
r(x) = \frac{ x^{-\frac{1}{3}}}{2} \left[ 1 + \frac{x \arctan\sqrt{x-1}}{\sqrt{x-1}}\right].
\label{RB}
\end{equation}
In the isotropic case $x = 1$, $r(1)=1$, $r^\prime(1)=0$, and Eq.~(\ref{epsilon2b}) is reduced to the standard equation of state used in \cite{Chojnacki:2007jc}.

The function $\Sigma$ appearing in Eq.~(\ref{enmomcon}) defines the entropy production. In the phenomenological approach \cite{Florkowski:2010cf} the following form of $\Sigma$ has been used
\begin{equation}
\Sigma(\sigma,x) = \frac{(1-\sqrt{x})^{2}}{\sqrt{x}}\frac{\sigma}{\tau_{\rm eq}},
\label{entropys}
\end{equation}
where the time-scale parameter $\tau_{\rm eq}$ controls the rate of equilibration. In Refs. \cite{Martinez:2010sc,Martinez:2010sd,Martinez:2012tu} a medium dependent $\tau_{\rm eq}$ was used, which was inversely proportional to the typical transverse momentum scale in the system.   In this case, as we will discuss in some detail below, the late time dynamics is that of second order viscous hydrodynamics.  If a constant value of $\tau_{\rm eq}$ is used, the system instead approaches perfect fluid behavior for $\tau \gg \tau_{\rm eq}$. 

If the generalized equation of state and the entropy production term are defined, the dynamical equations of anisotropic hydrodynamics become 5 equations for 5 unknown functions: three independent components of the fluid four-velocity $U^\mu$, the entropy density $\sigma$, and the anisotropy parameter $x$. In each case one has to additionally specify the value of the relaxation time or its dependence on other thermodynamics-like variables such as $\sigma$ or $x$.

\section{Microscopic approach}

In the microscopic approach \cite{Martinez:2010sc,Martinez:2010sd,Martinez:2012tu} one considers locally anisotropic momentum distributions of partons that are characterized by two different scales: $\lambda_\perp$ and $\lambda_\parallel$. They may be interpreted as the transverse and longitudinal temperatures ($LRF$ stands below for the local rest frame)
\begin{equation}
f_{LRF} = f \left( \frac{p_\perp}{\lambda_\perp},\frac{|p_\parallel|}{\lambda_\parallel}\right).
 \label{lambdas}
\end{equation}
In the studies of anisotropic quark-gluon plasma one often uses the Romatschke-Strickland form \cite{Romatschke:2003ms}
\begin{equation}
f_{LRF} = f \left( -\sqrt{\frac{p_\perp ^2}{\lambda_\perp  ^2} + 
\frac{p_\parallel^2}{\lambda_\parallel^2} } \,  \right) = f \left( - \frac{1}{\lambda_\perp}\sqrt{p_\perp ^2 + x \,\,p_\parallel^2 } \,  \right) ,
\label{RomStr}
\end{equation}
where $f$ is an arbitrary isotropic distribution function which is set by the underlying equilibrium distribution function for the particles being considered.
The covariant version of (\ref{RomStr}) reads
\begin{equation}
f= f \left( - \frac{1}{\lambda_\perp}\sqrt{(p\cdot U)^2 + \xi \,\,(p\cdot V)^2 } \,  \right),
\label{RomStrCov}
\end{equation}
where $\xi = 1 - x$. Using Eq.~(\ref{RomStrCov}) in the standard definitions of the energy-momentum tensor and the entropy flux one obtains the expressions for $T^{\mu\nu}$ and $\sigma^\mu$ of the form introduced in the previous Section.

In order to obtain the dynamical equations in the microscopic approach, one uses the Boltzmann equation. The zeroth moment of the Boltzmann equation, $p^\mu \partial_\mu f = C$, with the collision term  treated in the relaxation-time approximation, $C =  -p\cdot U \,\Gamma\, (f-f_{\rm eq})$, gives
\begin{equation}
\partial_\mu \left(\sigma U^\mu \right) =  \frac{1}{4} \int dP\ C = \Sigma 
= \frac{\Gamma}{4} \nonumber (n_{\rm eq}-n).
\label{zerothmom}
\end{equation}
Here $\Gamma$ is the inverse relaxation time ($\Gamma$ is inversely proportional to $\tau_{\rm eq}$ introduced earlier in the entropy source, however, the proportionality constant might be different from unity), $f_{\rm eq}$ is the background equilibrium distribution function, and $n$ is the parton density ($n_{\rm eq}$ is the corresponding equilibrium parton density). We note that for the Romatschke-Strickland form one finds $\sigma = 4 n$.

The first moment of the Boltzmann equation yields the energy-momentum conservation law
\begin{equation}
\partial_\mu \int dP\, p^\nu p^\mu  f =   \partial_\mu T^{\nu \mu} =0.  
\label{firstmom}
\end{equation}
These four equations are satisfied only if the collision term fulfills the condition
\begin{equation}
\int dP\, p^\nu C = - \int dP\, p\cdot U \, p^\nu \, \Gamma \, (f-f_{\rm eq}) =0.
\label{LMcon}
\end{equation}
The last equation is the Landau matching condition that specifies the temperature, $T$, appearing in the background equilibrium distributions $f_{\rm eq}$. If the Landau matching is used to eliminate $T$, we are left with Eqs.~(\ref{zerothmom}) and (\ref{firstmom}) which are 5 equations for 5 unknowns, similarly to the case described in the previous Section.

The structure of the equations of the anisotropic hydrodynamics obtained from the microscopic considerations is the same as that used in the phenomenological approach. On the other hand, the advantage of the microscopic approach is that the explicit form of the distribution function defines the entropy source term $\Sigma$ and the generalized equation of state. If the Romatschke-Strickland form is used, the generalized equation of state has the form (\ref{epsilon2b})--(\ref{PT2b}), however, the functions $\varepsilon_{\rm qgp}$ and $P_{\rm qgp}$ describe a massless parton gas (with Bose or Boltzmann statistics). The phenomenological and microscopic approaches have been compared in more detail in \cite{Ryblewski:2010ch}. 

Within the microscopic approach one can also perform an explicit matching to Israel-Stewart second order viscous hydrodynamics in the limit of small anisotropy.  In Ref.~\cite{Martinez:2010sc} it was shown that by linearizing the (0+1)-dimensional anisotropic hydrodynamics dynamical equations and requiring that they reproduce second-order viscous hydrodynamics, one can fix the relaxation rate $\Gamma$ appearing in Eq.~(\ref{zerothmom}) to be
\begin{eqnarray}
\Gamma &\equiv& \frac{2}{\tau_\pi}\,, \qquad \tau_\pi \equiv \frac{5}{4}\frac{\eta}{P_{\rm eq}}\, ,
\end{eqnarray}
\label{eq:hydromatch}
which for an ideal equation of state results in
\begin{equation}
\Gamma = \frac{2T}{5\bar\eta} \, ,
\label{eq:gammamatch}
\end{equation}
where $\bar\eta = \eta/S_{\rm eq}$ with $\eta$ being the shear viscosity, $S_{\rm eq}$ is the equilibrium entropy density,
and $T$ is the effective temperature determined via Landau matching in the first moment as described above.  Two points can be made using this result:  (1) the relaxation rate $\Gamma$ is proportional to the effective local temperature (fourth root of the energy density) and hence in regions of lower temperature, the relaxation to isotropy will be slower and (2) the relaxation rate $\Gamma$ is inversely proportional to the shear viscosity to entropy ratio and hence when the shear viscosity increases, one expects the relaxation to isotropy will be slower.  Both of these effects are important to understanding the dynamics of isotropization in the quark gluon plasma.

We want to emphasize that the microscopic approach presented here uses the zeroth and first moments of the Boltzmann equation. The zeroth moment in the standard approach to relativistic fluid dynamics is used to introduce the particle (or baryon) number conservation law. This is done by introducing a chemical potential $\mu$ together with the appropriate Landau matching condition which can be used to establish the value of $\mu$ (in a similar manner as the temperature was defined above). In our opinion this procedure is not appropriate for the systems produced in the early stages of relativistic heavy-ion collisions. Such systems consist primarily of gluons whose number is not conserved. In fact, the radiation of soft gluons leads to the increase of their numbers and to entropy production. Hence, the production of gluons is the main physical process which leads to equilibration of the system. Interestingly, such a physical picture is qualitatively similar to the bottom-up thermalization scenario \cite{Baier:2000sb}.

\section{Summary}

In this brief proceedings contribution we have attempted to review the motivation for and the foundational ideas behind anisotropic hydrodynamics.  The development of this framework has occurred using two separate methods, dubbed the phenomenological approach and the microscopic approach herein.  We have reviewed both methods pointing out the fundamental connections that exist between the two approaches.  In both cases presented, the framework of anisotropic hydrodynamics is constructed in such a way that the pressure (momentum) anisotropies of the system are built directly into the formalism. In our opinion this framework forms an alternative to the standard  viscous-hydrodynamics approach where one starts with the isotropic distribution and introduces potentially large corrections in the expansion.

The use of anisotropic hydrodynamics may help us to circumvent the early thermalization puzzle. It turns out that all soft hadronic observables may be successfully reproduced in the approach where the system is initially highly anisotropic \cite{Ryblewski:2012rr}. It turns out that the effects of the initial anisotropy may be compensated by a change of the initial conditions.  This points to a possible universality of the results for flow with respect to initial condition variation (at least with respect to the initial momentum-space anisotropy which is assumed).

There have been many other recent developments which we were not able to go into much detail about herein.  The approach presented has been generalized to incorporate the presence of dynamical longitudinal chromoelectric oscillations \cite{Florkowski:2012ax} and to include a mixture of two interacting anisotropic (quark and gluon) fluids which are characterized by different anisotropy parameters and equilibration times \cite{Florkowski:2012as}.  The work on mixtures of coupled quark and gluon fluids is potentially important because it demonstrates how to enforce number conservation in the baryonic (quark) sector while allowing number non-conservation in the gauge (gluon) sector.  An interesting project for the future would be to find more connections between anisotropic hydrodynamics and the studies of the anisotropic systems based on the AdS/CFT correspondance  \cite{Mateos:2011ix,Mateos:2011tv,Chernicoff:2012iq,Chernicoff:2012gu,Heller:2011ju,Heller:2012je}.

\newpage

\section*{Acknowledgments}

This work was supported by the Polish Ministry of Science and Higher Education under Grant No. N N202 263438 and the United States National Science Foundation under Grant No.~PHY-1068765.


\begin{thebibliography}{99}

\bibitem{Israel:1979wp} 
  W.~Israel and J.~M.~Stewart,
  Annals Phys.\  {\bf 118}, 341 (1979).
  
\bibitem{Muronga:2003ta} 
  A.~Muronga,
  Phys.\ Rev.\ C {\bf 69}, 034903 (2004)
  [nucl-th/0309055].
  
\bibitem{Baier:2006um} 
  R.~Baier, P.~Romatschke and U.~A.~Wiedemann,
  Phys.\ Rev.\ C {\bf 73}, 064903 (2006)
  [hep-ph/0602249].
  
\bibitem{Romatschke:2007mq} 
  P.~Romatschke and U.~Romatschke,
  Phys.\ Rev.\ Lett.\  {\bf 99}, 172301 (2007)
  [arXiv:0706.1522 [nucl-th]].
  
\bibitem{Dusling:2007gi} 
  K.~Dusling and D.~Teaney,
  Phys.\ Rev.\ C {\bf 77}, 034905 (2008)
  [arXiv:0710.5932 [nucl-th]].
  
\bibitem{Luzum:2008cw} 
  M.~Luzum and P.~Romatschke,
  Phys.\ Rev.\ C {\bf 78}, 034915 (2008)
  [Erratum-ibid.\ C {\bf 79}, 039903 (2009)]
  [arXiv:0804.4015 [nucl-th]].
  
\bibitem{Song:2008hj} 
  H.~Song and U.~W.~Heinz,
  J.\ Phys.\ G {\bf 36}, 064033 (2009)
  [arXiv:0812.4274 [nucl-th]].
  
\bibitem{Denicol:2010tr} 
  G.~S.~Denicol, T.~Kodama and T.~Koide,
  J.\ Phys.\ G {\bf 37}, 094040 (2010)
  [arXiv:1002.2394 [nucl-th]].
  
\bibitem{Schenke:2011tv} 
  B.~Schenke, S.~Jeon and C.~Gale,
  Phys.\ Lett.\ B {\bf 702}, 59 (2011)
  [arXiv:1102.0575 [hep-ph]].
  
\bibitem{Shen:2011eg} 
  C.~Shen, U.~Heinz, P.~Huovinen and H.~Song,
  Phys.\ Rev.\ C {\bf 84}, 044903 (2011)
  [arXiv:1105.3226 [nucl-th]].
  
\bibitem{Bozek:2011wa} 
  P.~Bozek,
  Phys.\ Lett.\ B {\bf 699}, 283 (2011)
  [arXiv:1101.1791 [nucl-th]].
  
\bibitem{Niemi:2011ix} 
  H.~Niemi, G.~S.~Denicol, P.~Huovinen, E.~Molnar and D.~H.~Rischke,
  Phys.\ Rev.\ Lett.\  {\bf 106}, 212302 (2011)
  [arXiv:1101.2442 [nucl-th]].
  
\bibitem{Bozek:2012qs} 
  P.~Bozek and I.~Wyskiel-Piekarska,
  Phys.\ Rev.\ C {\bf 85}, 064915 (2012)
  [arXiv:1203.6513 [nucl-th]].
  
\bibitem{Martinez:2009mf} 
  M.~Martinez and M.~Strickland,
  Phys.\ Rev.\ C {\bf 79}, 044903 (2009)
  [arXiv:0902.3834 [hep-ph]].
  

\bibitem{Kovchegov:2009he} 
  Y.~V.~Kovchegov,
  Nucl.\ Phys.\ A {\bf 830}, 395C (2009)
  [arXiv:0907.4938 [hep-ph]].
  
\bibitem{Lappi:2006hq} 
  T.~Lappi,
  Phys.\ Lett.\ B {\bf 643}, 11 (2006)
  [hep-ph/0606207].
  
\bibitem{Fukushima:2007ja} 
  K.~Fukushima,
  Phys.\ Rev.\ C {\bf 76}, 021902 (2007)
  [Erratum-ibid.\ C {\bf 77}, 029901 (2007)]
  [arXiv:0704.3625 [hep-ph]].
  
\bibitem{Kovchegov:2005ss} 
  Y.~V.~Kovchegov,
  Nucl.\ Phys.\ A {\bf 762}, 298 (2005)
  [hep-ph/0503038].
  
\bibitem{Krasnitz:2002mn} 
  A.~Krasnitz, Y.~Nara and R.~Venugopalan,
  Nucl.\ Phys.\ A {\bf 717}, 268 (2003)
  [hep-ph/0209269].
  
\bibitem{Florkowski:2010cf} 
  W.~Florkowski and R.~Ryblewski,
  Phys.\ Rev.\ C {\bf 83}, 034907 (2011)
  [arXiv:1007.0130 [nucl-th]].
  
\bibitem{Martinez:2010sc} 
  M.~Martinez and M.~Strickland,
  Nucl.\ Phys.\ A {\bf 848}, 183 (2010)
  [arXiv:1007.0889 [nucl-th]].
  
\bibitem{Ryblewski:2010bs}
  R.~Ryblewski and W.~Florkowski,
  J.\ Phys.\ G {\bf 38} (2011) 015104
  [arXiv:1007.4662 [nucl-th]].
  
\bibitem{Martinez:2010sd} 
  M.~Martinez and M.~Strickland,
  Nucl.\ Phys.\ A {\bf 856}, 68 (2011)
  [arXiv:1011.3056 [nucl-th]].
  
\bibitem{Ryblewski:2011aq} 
  R.~Ryblewski and W.~Florkowski,
  Eur.\ Phys.\ J.\ C {\bf 71}, 1761 (2011)
  [arXiv:1103.1260 [nucl-th]].
  
\bibitem{Martinez:2012tu} 
  M.~Martinez, R.~Ryblewski and M.~Strickland,
  Phys.\ Rev.\ C {\bf 85}, 064913 (2012)
  [arXiv:1204.1473 [nucl-th]].
  
\bibitem{Ryblewski:2012rr} 
  R.~Ryblewski and W.~Florkowski,
  Phys.\ Rev.\ C {\bf 85}, 064901 (2012)
  [arXiv:1204.2624 [nucl-th]].
  
\bibitem{Ryblewski:2012it} 
  R.~Ryblewski,
  arXiv:1207.0629 [nucl-th].
  
\bibitem{Chojnacki:2007jc} 
  M.~Chojnacki and W.~Florkowski,
  Acta Phys.\ Polon.\ B {\bf 38}, 3249 (2007)
  [nucl-th/0702030 [NUCL-TH]].

\bibitem{Rebhan:2008uj} 
  A.~Rebhan, M.~Strickland and M.~Attems,
  Phys.\ Rev.\ D {\bf 78}, 045023 (2008)
  [arXiv:0802.1714 [hep-ph]].
  
\bibitem{Martinez:2008di} 
  M.~Martinez and M.~Strickland,
  Phys.\ Rev.\ C {\bf 78}, 034917 (2008)
  [arXiv:0805.4552 [hep-ph]].

\bibitem{Romatschke:2003ms} 
  P.~Romatschke and M.~Strickland,
  Phys.\ Rev.\ D {\bf 68}, 036004 (2003)
  [hep-ph/0304092].
      
\bibitem{Florkowski:2012ax} 
  W.~Florkowski, R.~Ryblewski and M.~Strickland,
  Phys.\ Rev.\ D {\bf 86}, 085023 (2012)
  [arXiv:1207.0344 [hep-ph]].
  
\bibitem{Florkowski:2012as} 
  W.~Florkowski, R.~Maj, R.~Ryblewski and M.~Strickland,
  arXiv:1209.3671 [nucl-th].
  
\bibitem{Ryblewski:2010ch} 
  R.~Ryblewski and W.~Florkowski,
  Acta Phys.\ Polon.\ B {\bf 42}, 115 (2011)
  [arXiv:1011.6213 [nucl-th]].

\bibitem{Baier:2000sb} 
  R.~Baier, A.~H.~Mueller, D.~Schiff and D.~T.~Son,
  Phys.\ Lett.\ B {\bf 502}, 51 (2001)
  [hep-ph/0009237].
  
  \bibitem{Mateos:2011ix} 
  D.~Mateos and D.~Trancanelli,
  Phys.\ Rev.\ Lett.\  {\bf 107}, 101601 (2011)
  [arXiv:1105.3472 [hep-th]].
  
\bibitem{Mateos:2011tv} 
  D.~Mateos and D.~Trancanelli,
  JHEP {\bf 1107}, 054 (2011)
  [arXiv:1106.1637 [hep-th]].
  
  \bibitem{Chernicoff:2012iq} 
  M.~Chernicoff, D.~Fernandez, D.~Mateos and D.~Trancanelli,
  JHEP {\bf 1208}, 100 (2012)
  [arXiv:1202.3696 [hep-th]].
  
\bibitem{Chernicoff:2012gu} 
  M.~Chernicoff, D.~Fernandez, D.~Mateos and D.~Trancanelli,
  JHEP {\bf 1208}, 041 (2012)
  [arXiv:1203.0561 [hep-th]].
  
\bibitem{Heller:2011ju} 
  M.~P.~Heller, R.~A.~Janik and P.~Witaszczyk,
  Phys.\ Rev.\ Lett.\  {\bf 108}, 201602 (2012)
  [arXiv:1103.3452 [hep-th]].
  
\bibitem{Heller:2012je} 
  M.~P.~Heller, R.~A.~Janik and P.~Witaszczyk,
  Phys.\ Rev.\ D {\bf 85}, 126002 (2012)
  [arXiv:1203.0755 [hep-th]].
  
\end{thebibliography}
\end{document}